\documentclass[superscriptaddress,prl,twocolumn,nofootinbib]{revtex4-1}
\usepackage{pdfpages}
\usepackage{epstopdf}
\usepackage{color}
\usepackage{graphicx}
\usepackage[justification=centerlast]{caption}
\usepackage{subcaption}
\usepackage{amsmath}
\newcommand{\ket}[1]{|#1\rangle}

\usepackage{hyperref} % has to be last command in preamble to work right
\hypersetup{bookmarks=true,colorlinks=true,linkcolor=blue,citecolor=blue}

\begin{document}

\title{
Chip-based Quantum Key Distribution
}

\author{P. Sibson}
\email[]{philip.sibson@bristol.ac.uk}
\author{C. Erven}
\author{M. Godfrey}
\affiliation{Centre for Quantum Photonics, H. H. Wills Physics Laboratory and Department of Electrical and Electronic Engineering, University of Bristol, Merchant Venturers Building,  Woodland Road, Bristol BS8 1UB, UK.}
\author{S. Miki}
\author{T. Yamashita}
\affiliation{National Institute of Information and Communications Technology (NICT), 588-2 Iwaoka, Kobe 651-2492, Japan}
\author{M. Fujiwara}
\author{M. Sasaki}
\affiliation{National Institute of Information and Communications Technology (NICT), 4-2-1 Nukui-Kitamachi, Koganei, Tokyo 184-8795, Japan}
\author{H. Terai}
\affiliation{National Institute of Information and Communications Technology (NICT), 588-2 Iwaoka, Kobe 651-2492, Japan}
\author{M. G. Tanner}
\author{C. M. Natarajan}
\author{R. H. Hadfield}
\affiliation{School of Engineering, University of Glasgow, G12 8QQ, United Kingdom}
\author{J. L. O'Brien}
\author{M. G. Thompson}
\email[]{mark.thompson@bristol.ac.uk}
\affiliation{Centre for Quantum Photonics, H. H. Wills Physics Laboratory and Department of Electrical and Electronic Engineering, University of Bristol, Merchant Venturers Building,  Woodland Road, Bristol BS8 1UB, UK.}

%===============================================================================
% Abstract
%===============================================================================
\begin{abstract}
\noindent Improvement in secure transmission of information is an urgent practical need for governments, corporations and individuals. Quantum key distribution \cite{QKDREVIEW,Scarani2009} (QKD) promises security based on the laws of physics and has rapidly grown from proof-of-concept to robust demonstrations \cite{Yoshino:13,Korzh:13,Dixon:15,sasaki2015quantum} and even deployment of commercial systems \cite{idQuantique,MagiQ,Quintessence}. 
Despite these advances, QKD has not been widely adopted, and practical large-scale deployment will likely require integrated chip-based devices for improved performance, miniaturisation and enhanced functionality, fully integrated into classical communication networks. 
Here we report low error rate, GHz clocked QKD operation of an InP transmitter chip and a SiO$_x$N$_y$ receiver chip---monolithically integrated devices that use state-of-the-art components and manufacturing processes from the telecom industry.	We use the reconfigurability of these devices to demonstrate three important QKD protocols---BB84, Coherent One Way (COW) and Differential Phase Shift (DPS)---with performance comparable to state-of-the-art. These devices, when combined with integrated single photon detectors, satisfy the requirements at each of the levels of future QKD networks---from point-of-use through to backbones---and open the way to operation in existing and emerging classical communication networks.
\end{abstract}

\maketitle

%===============================================================================
% Introduction
%===============================================================================
\noindent Delivery of the promise of QKD will require moving to an integrated network, where, as with the classical communication infrastructure, each level has particular requirements that must be met. Use of QKD in hand-held and field deployable devices, or to secure the \textit{`Internet of Things'} for example, will require devices with a small footprint and high robustness to environmental conditions. `QKD-to-the-home' will require moderate bandwidth, low cost devices, and co-existence with emerging fibre-to-the-home transceivers. Backbones for QKD metro and long-haul networks will require compatibility, and in some cases direct integration, with photonic and electronic semiconductor devices and systems. Also crucial will be the capability to reconfigure QKD protocols on the fly and overcome the bottlenecks currently limiting long-range, high-speed key distribution. Use of common devices will enable unification across all levels of the network. Ultimately, a QKD network must be seamlessly integrated with the classical communication network with the distinction between classical and quantum operation ideally being defined in software not hardware.

While extreme levels of integration have been achieved in the microelectronics industry over the past decades, it is only recently that size, cost and power consumption considerations have demanded higher levels of integration in photonics. Fibre-to-the-home, data centre, and 100~Gbs$^{-1}$ metro and long-haul network applications have driven the development of the InP platform to the point of full integration of laser sources, amplifiers, modulators and detectors \cite{INPREVIEW}. Integrated photonics \cite{INTREVIEW} is thus poised to deliver major benefits to QKD technology and networks \cite{DARPA,SECOQC,TOKYO} by allowing the miniaturisation of components and circuits for hand-held and field deployable devices. It also provides highly robust manufacturing processes which help reduce cost for personal devices. Finally, the complexity achievable with the integrated platform enables practical implementation of mutli-protocol operation for flexility, multiplexing for higher rates, and additional monitoring and certification circuits to protect against side-channel attacks \cite{QKDREVIEW} in a fibre network.

\begin{figure*}[t!]
	\begin{center}
	   \includegraphics[trim = 0mm 10mm 0mm 10mm, width=1\textwidth]{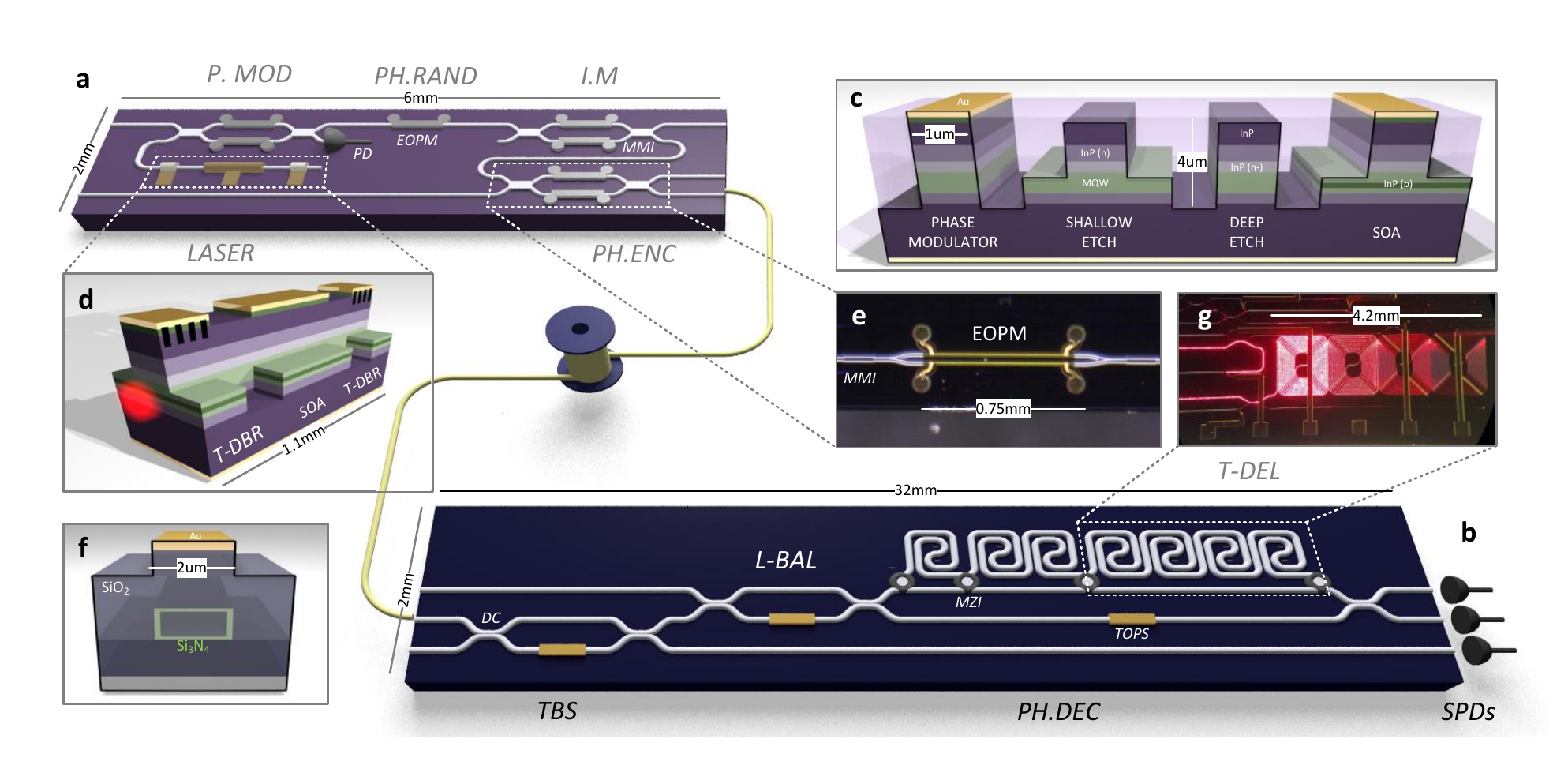}
	\end{center}
    \caption{Integrated photonic devices for quantum key distribution consisting of: (a) A monolithically integrated Indium Phosphide (InP) transmitter for GHz clock rate, reconfigurable, multi-protocol QKD. (b) A Silicon Oxynitride (Triplex) photonic receiver circuit for reconfigurable, multi-protocol QKD that passively decodes the quantum information with off-chip single photon detectors. (c) The  InP technology platform waveguide cross-section \cite{INPREVIEW}. (d) Wavelength tunable continuous-wave laser, formed from two tuneable Distributed Bragg Reflectors (T-DBR) and a semiconductor optical amplifier (SOA). (e) Microscopic image of electro-optic phase modulators in Mach-Zehnder interferometer. (f) The SiO$_x$N$_y$ Triplex waveguide cross-section, with metalisation for heating elements \cite{Triplexconf}. (g) Microscopic image of the receiver delay lines.}
	\label{fig:CHIP}
\end{figure*}

While there have been individual demonstrations of time-bin decoding \cite{tanaka2012high}, miniaturisation \cite{RFI}, and reconfigurability \cite{Korzh:13} in integrated devices; here we report QKD operation of complex devices that meet the requirements outlined above. We used the InP platform to implement a monolithically integrated  transmitter (Figure~\ref{fig:CHIP}a), consisting of a tunable laser, optical interferometers, electro-optic phase modulators and a PIN photodiode. We implement a receiver (Figure~\ref{fig:CHIP}b), consisting of a photonic circuit with thermo-optic phase shifters and reconfigurable delay line in the SiO$_x$N$_y$ platform, and off-chip single photon detectors. Both photonic systems were manufactured using state-of-the-art industrial fabrication processes (Oclaro and LioniX, respectively) and were designed for multi-protocol reconfigurable operation. We show performance of the photonic devices with clock rates up to 1.7~GHz, a quantum bit error rate (QBER) as low as 0.88\%, and estimated secret key rates up to 568~kbs$^{-1}$, for an emulated 20~km fibre link. Together with the development of integrated single photon detectors\cite{hadfield2009single}, these devices point the way to seamless integration with existing and emerging classical communication systems, including microelectronics.

%===============================================================================
% Body
%===============================================================================

Figure 1 shows a schematic of the chip-to-chip QKD system. For the transmitter device, the InP material system was chosen to meet the requirements of fast active electro-optics (with GHz operating speeds) and monolithic integration with the laser source.
For the receiver device, the SiO$_x$N$_y$ material system was chosen to minimise photon loss from fibre-to-chip coupling and waveguide propagation loss, whilst maintaining a compact footprint. Both devices, along with fibre coupled single photon detectors, represent the full photonic QKD system.

The InP-based transmitter chip was fabricated using an advanced active-passive integration technology \cite{INPREVIEW}, where a multistep epitaxial growth process provides large flexibility in the waveguide structure (see Figure~\ref{fig:CHIP}(c)). The on-chip tunable laser (Figure~\ref{fig:CHIP}(d)) was formed from two distributed Bragg reflectors (DBR) and a semiconductor optical amplifier (SOA). When operated in continuous wave (CW) the laser source exhibited single mode behaviour with FWHM of 34~pm, a side-mode suppression ratio of $>$50~dB, and an operating wavelength of 1550~nm with $\sim$10~nm tuning range. Short electrical pulses applied to the reverse biased electro-optic phase modulator (EOPM) in the first MZI enabled optical pulse generation with $<$150~ps duration and $\sim$30~dB extinction ratio. The exact timing between consecutive pulses could be accurately controlled by the driving electronics (see SI), and the on-chip photodiode was used to monitor the laser intensity and provide feedback to stabilise the laser current. The remaining electro-optic phase modulators and MZIs were used to drive the different QKD protocols and to attenuate the laser pulses to the single photon level. Light was coupled out of the device using a lensed optical fibre.

The SiO$_x$N$_y$ receiver chip was fabricated using the TripleX technology platform \cite{Triplexconf}, where alternating layers of Si$_3$N$_4$ and SiO$_2$ were deposited and etched to create a hollow box structure to guide light in a high index-contrast but low loss waveguides ($\sim$0.5~dB/cm), and with low coupling loss between chip and fibre ($\sim$2~dB), yielding a total loss $\sim$ 9dB for BB84 configuration. Metal layers on top of the structure created thermo-optic phase shifters for circuit reconfigurability.
The first MZI acts as a tunable beamsplitter (TBS) and taps off a portion of the incoming signal, which was routed to a single photon detector and used primarily for the COW protocol. The second MZI (L-BAL) acts to balance the losses in the asymmetric MZI (AMZI), which incorporates a digitally reconfigurable delay line, tunable from 0 to 2.1~ns in steps of 300~ps. The thermo-optic phase shifter (TOPS) within the AMZI was used calibrate the phase relationship between the two arms of the interferometer. Light was coupled out of the device and into external fibre coupled superconducting nanowire single-photon detectors mounted in a closed cycle refrigerator\cite{SNSPDS} which had an system detectorion efficiency of $\sim$45\% from the fibre input, a temporal jitter of $\sim$50~ps, and a dead-time of $\sim$10~ns.

The highly reconfigurable nature of the transmitter and receiver devices allowed the implementation of a number of different QKD protocols. Here, we specifically investigated the three protocols of BB84, COW and DPS.

\begin{figure}[htbp]
    \centering
        \includegraphics[trim = 0mm 15mm 0mm 5mm, width=1.05\columnwidth]{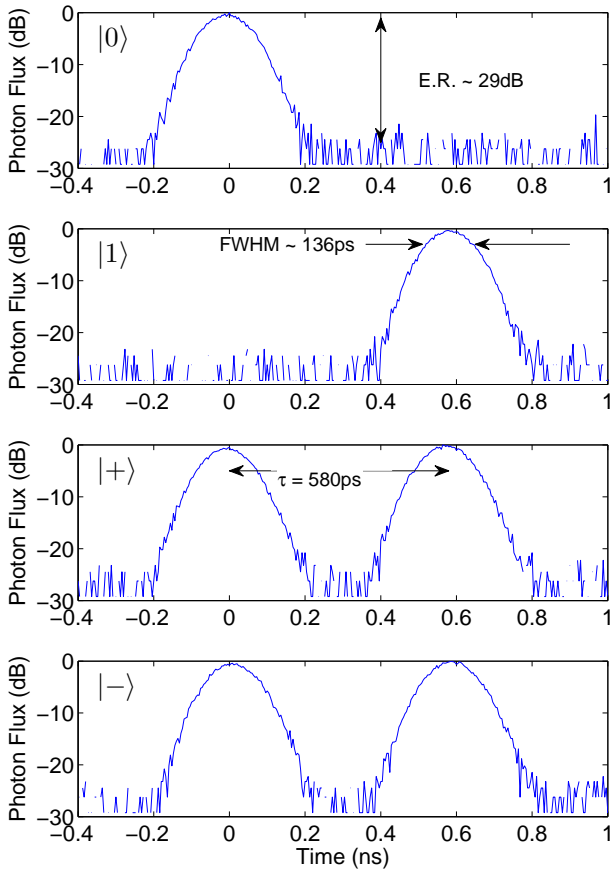}
    \caption{Transmitter output for the four BB84 states, demonstrating the 136~ps FWHM pulses with near 30~dB extinction, and temporal separation of 580~ps.}
    \label{fig:LASERS}
\end{figure}

\begin{figure*}[t!]
	\begin{center}
		\includegraphics[trim = 50mm 10mm 0mm 0mm, width=1.1\textwidth]{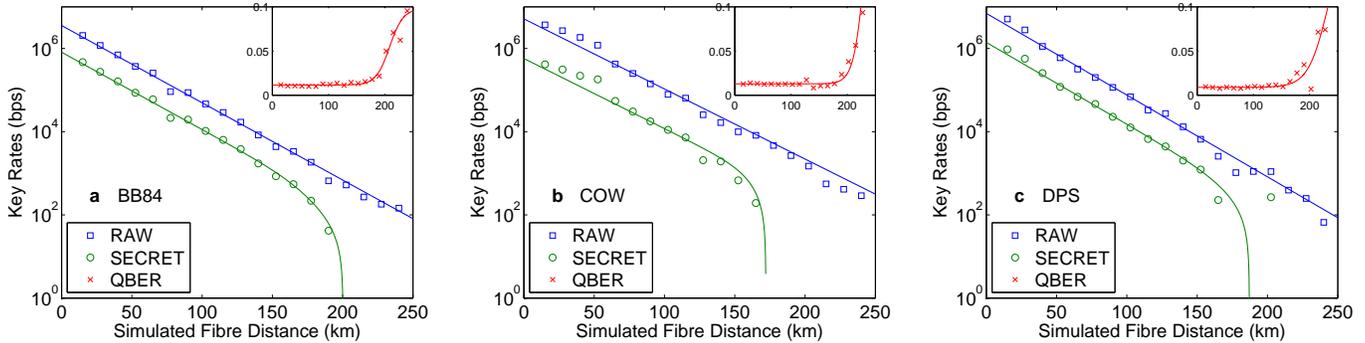}
	\end{center}
    \caption{Experimental results for (a) BB84, (b) COW, and (c) DPS showing the raw detection rate, estimated secret key rate, and relevant QBER. For BB84 the QBER is derived from the timing and phase errors, while for COW the QBER is derived from the timing error and security of the channel is estimated from phase coherence between successive pulses, and finally for DPS the QBER is estimated based on the error from the phase encoded information. State (or clock) rates of 560~MHz, 860~MHz, and 1.76~GHz were used for BB84, COW, and DPS respectively.}
	\label{fig:RATES}
\end{figure*}

The BB84 \cite{BB84} QKD protocol was implement using time-bin encoding, where $\ket{0}$ was encoded by a photon in the first time bin and $\ket{1}$ was encoded by a photon in the second time bin, while $\ket{+}$ was encoded by a photon in a superposition of the first and second time-bin with zero relative phase, and $\ket{-}$ was encoded by a photon in a superposition of being in the first and second time bin with a $\pi$ relative phase, as illustrated in Figure~\ref{fig:LASERS}.
The BB84 protocol transmits one of two orthogonal states chosen at random, encoded in one of two randomly chosen non-orthogonal bases. We used the Z-basis $\{ \ket{0}, \ket{1} \}$ and the X-basis $\{ \ket{+}, \ket{-} \}$.

The CW laser source was modulated (P.MOD) to select the time bin choice, which were then phase randomised with a single electro-optic modulator (PH.RAND) before being attenuated and intensity modulated (I.M). The intensity of the $\{ \ket{+},  \ket{-} \}$ states was reduced by half, compared to the $\{ \ket{0}, \ket{1} \}$ states in order to maintain the same average photon number per state. The intensity modulator was also used to encode the decoy photon levels required to mitigate multi-photon contamination for security \cite{2005Decoyproof}. The final MZI encoded the relative phase between successive time bins to implement the $\ket{-}$ state.

Within the receiver chip, the digitally tunable delay line was reconfigured to match the 600ps time interval between time-bins from the transmitter device. The phase decoding AMZI overlapped successive time-bins creating three possible time-slots within which to detect photons. Phase information interfered in the middle time-slot allowing measurements in the $\{ \ket{+}, \ket{-} \}$ basis, whereas time of arrival information in the first and third time-slots measured in the $\{ \ket{0}, \ket{1} \}$ basis.
This decoding allows for a passive optical circuit, with the detection event constituting the random basis choice, thus removing the requirement for GHz rate active elements and quantum random number generators in the receiver.

The COW protocol \cite{stucki2005fast} transmits pulses in pairs, encoding $\ket{0}$ with the first bin and $\ket{1}$ with the second. Again the pulse modulated CW laser was used to generate pulses in these time-bins. While the key was generated unambiguously from the time of arrival of the single photon in a pair, security of the channel was determined by measuring the visibility from interfering successive photon pulses. A decoy state, with photon pulses in each time-bin ($\ket{0}$ and $\ket{1}$), was included to increase the probability of occupied successive pulses, allowing a more accurate measurement of interference. Using the first MZI, the receiver routes a larger proportion of the input signal to single photon detectors for key generation, and a smaller proportion to the AMZI for visibility measurement.

Finally, the DPS protocol \cite{inoue2002differential} encodes information within the relative phase, 0 and $\pi$, of a train of photon pulses generated from the temporally modulated CW laser. The information was decoded unambiguously through the AMZI by interfering successive pulses, providing a QBER based on the number of incorrect counts at the wrong output of the phase decoding circuit. The security of the channel was determined by bounding the possible information an adversary could extract, that in turn would cause errors in the transmitted information.

%===============================================================================
% Results
%===============================================================================

\begin{table}[t]
	\centering
	\begin{tabular}{|c|c|c|c|c|c|c|}
		\hline Protocol & $\mu$ & State & QBER & QBER &  Raw & Secret \\
		  & (per & Rate & Time & Phase &  Rate & Rate \\
		  & pulse) & (GHz) & (\%) & (\%) &  (Mbps) & (kbps) \\
		\hline BB84 & 0.45 & 0.56 & 1.17 & 0.92 & 1.51 & 345  \\
		\hline COW & 0.28 & 0.86 & 1.37 & 1.36 & 2.67 & 311 \\
		\hline DPS & 0.28 & 1.76 & - & 0.88 & 2.78 & 565 \\
		\hline
	\end{tabular}
    \caption{Comparison of parameters and measured rates for the three QKD protocols over an emulated fibre link of 20~km, assuming 0.2~dB/km, using a digital variable attenuator.}
	\label{tab:protocols}
\end{table}

Each of the above three protocols was implemented on the chip-to-chip system, where the length of optical fibre link was emulated using a variable optical attenuator to induce channel loss, where a loss of 0.2~dB/km was assumed (standard within telecommunications fibres at 1550~nm), although rates could be improved through use of low loss fibres \cite{2015longrangecow}, and optimising the SNSPDs for ultra low dark counts \cite{2014Shibata72db}. The effects of dispersion were considered negligible for the broad $\sim$150~ps pulses used here. The performance of our integrated devices for all three protocols is shown in Figure~\ref{fig:RATES}, where the raw key rate, estimated secret key rate, and QBER observed are plotted. For BB84, using an attenuation equal to 20~km of fibre we obtained an estimated secret key rate of 345~kbits/s using a clock rate of 560~MHz; using average single photon numbers of 0.45, 0.1, and 5.0$\times10^{-4}$ for the signal and two decoy states chosen with probabilities of 0.8, 0.15, and 0.05 respectively; and observed an average QBER of 1.05\%. The secret key rate for BB84 was calculated using the raw and sifted key rates, and the measured QBER, using the security proof of Ma \emph{et al.}\cite{2005Decoyproof}.

For COW, again using an attenuation equal to 20~km of fibre we obtained an estimated secret key rate of 311~kbits/s using a clock rate of 0.86~GHz with a QBER of 1.37\% due to timing information and a QBER of 1.36\% due to the interferometer and security of the channel. The secret key rate of COW was calculated using the sifted key rate and measured visibilities according to the security proof by Branciard \emph{et al.}\cite{branciard2008upper} shown to be a secure upper bound for collective attacks.

For DPS at the same attenuation we obtained an estimated secret key rate of 565~kbits/s using a clock rate of 1.76~GHz and measuring a QBER of 0.88\%. The secret key rate of DPS was calculated by measuring the key errors and visibilities according to the security proof by Branciard \emph{et al.}\cite{branciard2008upper} and is limited to collective attacks.

%===============================================================================
% Discussion/Summary
%===============================================================================

A summary of these results are presented in Table 1, where in all cases we show a performance comparable to the state-of-the-art in current fibre and bulk optical systems\cite{QKDREVIEW}. This work demonstrates the feasibility of using fully integrated devices within QKD systems, implementing three important protocols by utilising the reconfigurability of the devices. The integrated photonic platform allowed us to demonstrate miniaturised devices exploiting robust, low-cost manufacturing processes, that allow flexibility in fibre network settings.

These devices could be readily adapted to implement more protocols, such as the reference-frame independent and measurement-device independent QKD protocols \cite{RFIQKD,MDIQKD}. Also the tunablity of the laser source enables flexibility in the wavelength of operation, this combined with the complexity achievable with the platform will be key to enabling high capacity wavelength division multiplexing schemes of the quantum channel in a practical implementation. The increased complexity allowed by integrated photonics will facilitate the implementation of further monitoring and certification circuits, protecting against security flaws and side channel attacks\cite{QKDREVIEW} with minimal change in footprint and cost.

Compatibility with current integrated photonic telecommunication hardware will ultimately allow seamless operation alongside classical communications transceivers, enabling hybrid classical and quantum communications devices. Moreover, the ability to scale up these integrated circuits to 100's or even 1000's of components \cite{INPREVIEW} opens the way to new and advanced integrated quantum communications technologies.

%===============
% Acknowledgements
%===============
%\begin{acknowledgments}
%\textbf{Author Contributions:}

%==========
% Bibliography
%==========
\bibliographystyle{ieeetr}
%\bibliography{QKDBIB}

\end{document}